\def\tc{t_{\rm cool}}
\def\tff{t_{\rm ff}}
\def\tctff{\tc / \tff}

\documentclass{emulateapj}

\begin{document}

\slugcomment{\apj\ Letters, submitted 13 May 2015 (Printed \today)}
\title{Precipitation-Regulated Star Formation in Galaxies}
\author{G. Mark Voit\altaffilmark{1,2}, Greg L. Bryan\altaffilmark{3}, Brian W. O'Shea\altaffilmark{1}, Megan Donahue\altaffilmark{1},  
         } 
\altaffiltext{1}{Department of Physics and Astronomy,
                 Michigan State University,
                 East Lansing, MI 48824} 
\altaffiltext{2}{voit@pa.msu.edu}
\altaffiltext{3}{Department of Astronomy,
                 Columbia University,
                 New York, NY} 
                           
\begin{abstract}
Galaxy growth depends critically on the interplay between radiative cooling of cosmic gas and the resulting energetic feedback that cooling triggers.   This interplay has proven exceedingly difficult to model, even with large supercomputer simulations, because of its complexity.  Nevertheless, real galaxies are observed to obey simple scaling relations among their primary observable characteristics.  Here we show that a generic emergent property of the interplay between cooling and feedback can explain the observed scaling relationships between a galaxy's stellar mass, its total mass, and its chemical enrichment level, as well as the relationship between the average orbital velocity of its stars and the mass of its central black hole. These relationships naturally result from any feedback mechanism that strongly heats a galaxy's circumgalactic gas in response to precipitation of colder clouds out of that gas, because feedback then suspends the gas in a marginally precipitating state.
\end{abstract}


\setcounter{footnote}{0}

\section{Introduction}
\label{sec-Intro}

Radiative cooling is essential to galaxy formation because shock waves heat cosmic gas to million-degree temperatures as gravity pulls it into galaxies. Early models of galaxy formation therefore focused on comparing the cooling time $\tc$ of the gas with the free-fall time $\tff = (2r/g)^{1/2}$ required to fall from radius $r$ at acceleration $g$ \citep{ReesOstriker1977MNRAS.179..541R,Silk1977ApJ...211..638S,Binney1977ApJ...215..483B,FallRees1985ApJ...298...18F}.  Gas with $\tc  < \tff$ was presumed to become thermally unstable, leading to the formation of a multiphase medium consisting of cold star-forming clouds embedded in an ambient million-degree atmosphere.  Later models recognized that uncompensated cooling would produce many more stars than are observed, requiring a feedback response that inhibits cooling \citep{wr78,DekelSilk1986ApJ...303...39D,WhiteFrenk1991ApJ...379...52W}.  Supernovae can provide sufficient feedback in small galaxies but cannot fully compensate for cooling in larger ones \citep{Benson+2003ApJ...599...38B,Croton+2006MNRAS.365...11C,Bower+2006MNRAS.370..645B}.  Instead, accretion of cooling gas onto a central supermassive black hole seems required to generate the feedback that regulates star formation in the universe's largest galaxies \citep{TaborBinney1993MNRAS.263..323T,SilkRees1998AA...331L...1S,McNamaraNulsen2012NJPh...14e5023M}.  

Exactly how black-hole feedback becomes tuned to compensate for radiative cooling at distances $\gtrsim 10^4$~parsecs has been a resilient mystery, but there have been some recent breakthroughs.  It now appears that the black-hole fueling rate in giant galaxies depends sensitively on development of a multiphase medium \citep{ps05,Voit+08,Cavagnolo+08}.  Numerical simulations indicate that circumgalactic gas in approximate thermal balance develops a multiphase medium when it becomes thermally unstable \citep{ps10,McCourt+2012MNRAS.419.3319M,Sharma+2012MNRAS.420.3174S}.  In simulated galaxies, cold clouds precipitate out of the hot medium when $\tc$ drops below $\sim 10 \tff$ and then accrete chaotically onto the galaxy's central black hole, fueling a strong feedback response \citep{Gaspari+2012ApJ...746...94G,Gaspari+2013MNRAS.432.3401G,Gaspari+2014arXiv1407.7531G,LiBryan2014ApJ...789...54L,LiBryan2014ApJ...789..153L}. This jolt of heating subsequently expands the circumgalactic medium, raising $\tc$ and diminishing the precipitation.  Interplay between precipitation and feedback therefore causes $\min ( \tctff )$ to fluctuate in the range $5 \lesssim \min ( \tctff ) \lesssim 20$ and to settle near $\tctff \approx 10$.

X-ray observations of circumgalactic gas around giant galaxies are revealing a distinct cooling-time floor at $\tc \approx 10 \tff$ \citep{VoitDonahue2015ApJ...799L...1V,Voit+2015Natur.519..203V,Voit+2015ApJ...803L..21V}, confirming a key prediction of those numerical models \citep{Sharma+2012MNRAS.427.1219S}.  Here we explore the implications for galaxy evolution if feedback driven by precipitation of cold clouds produces a similar cooling-time floor in galactic systems of all sizes, from large galaxy clusters all the way down to the smallest dwarf galaxies, and we show that this simple hypothesis can account for many of the observed properties of galaxies.

\section{The Precipitation Threshold}
\label{sec-threshold}

There is not yet a complete theory for the origin of the critical $\tctff$ value. In lieu of a complete theory, we follow the observations and assume that $\tctff \approx 10$ is the threshold at which precipitation triggers a strong feedback response that reheats the circumgalactic gas.  If the amount of precipitating gas rapidly increases as $\tctff$ drops, then any form of feedback fueled by precipitation causes the system to linger near $\tctff \approx 10$, regardless of whether the feedback comes from black-hole accretion or supernova explosions.  The electron number density at radius $r$ in a generic precipitation-regulated system is therefore
\begin{equation}
  n_e(r) \approx \frac {3kT} {10 \Lambda \tff} \left( \frac {n_e + n_i} {2 n_i} \right) \; \; ,
\end{equation}
where $T$ is gas temperature, $n_i$ is ion number density, and $\Lambda$ is the usual cooling funtion \citep{Sharma+2012MNRAS.427.1219S}.  In such a system, the radius $r_{500}$ that encompasses an average mass density 500 times the critical density is an approximate upper limit on the size of the precipitating region, because $\tc \approx H_0^{-1}$ at that radius.  Intriguingly, spectroscopic absorption-line studies of the circumgalactic medium indicate that small clouds of low-ionization gas are common inside of $r_{500}$ but are rare outside of that radius \citep{LiangChen2014MNRAS.445.2061L}, suggesting that the low-ionization material might be incipient precipitation.

Condensation and precipitation of the circumgalactic medium supply the cold gas needed for star formation in a precipitation-regulated galaxy.  One expects that gas supply to be proportional to the integral of $n_e \tc^{-1}$ within the volume bounded by $r_{500}$. We therefore express the time-averaged star formation rate in a precipitation-regulated galaxy as
\begin{equation}
  \dot{M}_* \approx \left(  \frac {3 \pi} {50} \frac {G \mu_e m_p kT} {f_b \Lambda} \right) \epsilon_* f_{\rm b} M_{500} \; \; ,
  \label{eq:mdotstar}
\end{equation}
where $\epsilon_*$  is an efficiency parameter accounting for the complexities of star formation and feedback, $f_{\rm b}$ is the cosmic baryon mass fraction, $\mu_e m_p$ is the mean mass per electron, $M_{500}$ is the total mass within $r_{500}$, and the approximation assumes that $T$ and $r/\tff$ are independent of radius.  (In giant galaxies, the limiting value of $\dot{M}_*$ is smaller because the cosmological upper limit on $n_e$ is more stringent at large radii than the precipitation threshold.)

According to equation (1), feedback in a precipitation-regulated galaxy limits cooling by driving down the density of its circumgalactic medium.  It does not permanently expel all the circumgalactic gas from the system but rather inflates it like a balloon, pushing gas that would have been inside of $r_{500}$ to larger radii, thereby lowering the mean density of the precipitating region.  Electron densities around dwarf galaxies, which have comparatively low $T$ and large $\Lambda$, are consequently several orders of magnitude smaller than those around giant galaxies.  

According to equation (2), the average star-formation rate of a precipitation-regulated galaxy depends inversely on chemical enrichment of the circumgalactic gas through the cooling function $\Lambda$, which is proportional to the abundances of elements heavier than H and He for gas in the $10^5-10^7$~K temperature range.  This inverse dependence arises because enrichment makes the circumgalactic medium more volatile.  Paradoxically, adding heavy elements to a precipitating system reduces $\dot{M}_*$ because feedback is triggered at a lower circumgalactic gas density.  Observations of real star-forming galaxies show that they share this qualitative feature, in that their star-formation rates anticorrelate with chemical enrichment \citep{Ellison+2008ApJ...672L.107E,Mannucci+2010MNRAS.408.2115M}.

\section{Saturation of Enrichment}
\label{sec-saturation}

Coupling between chemical enrichment and star formation in a precipitation-regulated galaxy causes its enrichment level to saturate.  In order to illustrate how saturation arises, we adopt a simplistic assumption: that the heavy elements made by stars mix evenly with all the gas associated with a galaxy, including its circumgalactic medium.  The mass $M_{Z,{\rm  gas}}$ of heavy elements in the gas then accumulates according to $\dot{M}_{Z,{\rm gas}} = (Y - Z_{\rm gas} ) \dot{M}_*$, where $Y$ is the heavy-element yield from a single generation of star formation and $Z_{\rm gas} = M_{Z,{\rm gas}} /M_{\rm gas}$.  Note that $M_{\rm gas} = f_{\rm b} M_{500} - M_*$ represents the entire gas mass originally associated with the galaxy's dark matter, less the amount that has turned into stars, and generally exceeds the gas mass within $r_{500}$.

This chemical enrichment model is oversimplified but nicely clarifies the mechanism of saturation. Taking the derivative of $Z_{\rm gas}$ with respect to time, we find
\begin{equation}
  \dot{Z}_{\rm gas} = \left[ Y \dot{M}_* - Z_{\rm gas} f_{\rm b} \dot{M}_{500} \right] \frac {1} {M_{\rm gas}} \; \; .
\end{equation}
The first term in brackets is a source term for element production, and the second is a dilution term accounting for inflowing unenriched gas.  Together they tune both the galaxy's heavy-element abundance and its star-formation rate to particular values determined by the galaxy's gravitational potential well.  Early in a galaxy's history, when enrichment is low, the source term dominates and $Z_{\rm gas}$ increases. However, enrichment also increases the ability of circumgalactic gas to cool, which promotes precipitation.  Feedback therefore causes the circumgalactic gas to expand until it reaches the precipitation threshold at $\tctff \approx 10$. In this state of marginal stability, expansion continues while $Z_{\rm gas}$ increases, which drives down the star-formation rate. As a result, the heavy-element abundance of a precipitation-regulated galaxy plateaus when it reaches saturation at $Z \approx Y \dot{M}_* ( f_{\rm b}  \dot{M}_{500} )^{-1}$. 

Chemical enrichment of real galaxies is undoubtedly more complicated.  However, heavy-element abundances in any precipitation-regulated system will still saturate as long as enrichment of the circumgalactic medium is proportional to the total production of elements by the galaxy's stars.  In such a system, chemical enrichment of the circumgalactic medium reduces its condensation rate at the precipitation threshold, which is what is required for saturation to happen. 

\section{Galactic Star-Formation Histories} 
\label{sec-SFR}

Approximate star-formation histories for precipitation-regulated galaxies can be obtained by rewriting equation (\ref{eq:mdotstar}) as
\begin{equation}
  H^2 \frac {d} {dt} M_*^2 \, \approx \, (2 \times 10^{11} \, M_\odot )^2 \left(  \frac {Z_\odot} {Y} \right) \epsilon_* \,
  					H_0^3 \, v_{200}^{10} \; \; ,
\label{eq:dM2dt}
\end{equation}
where $H$ is the Hubble expansion parameter and $v_{200}$ is the circular velocity $v_c \approx (GMr^{-1})^{1/2}$ in units of $200 \, {\rm km \, s^{-1}}$.  To get equation~(\ref{eq:dM2dt}), we assumed $\Lambda \approx (1.8 \times 10^{-22} \, {\rm erg \, cm^3 \, s^{-1}}) (Z/Z_\odot) (T/ 10^6 \, {\rm K})^{-1}$, which is within 50\% of the \citet{sd93} tabulation for $10^5 \, {\rm K} < T < 10^7 \, {\rm K}$. We also assumed $Z \approx Y M_* / (f_{\rm b} M_{500})$ and $kT \approx \mu m_p v_c^2$, as appropriate for hydrostatic gas with $n_e \propto r^{-1}$ in an isothermal potential.  Mass growth of a typical galactic halo follows $\dot{M}_{500} = \beta M_{500} t^{-1}$, with $\beta \approx 0.5 (H_0 t) (1+z)^{2.35}$ in a $\Lambda$CDM cosmology \citep{NeisteinDekel2008MNRAS.388.1792N}, and using this approximation to integrate equation~(\ref{eq:dM2dt}) gives
\begin{equation}
  M_* \approx (2 \times 10^{11} \, M_\odot ) \left(  \epsilon_* \frac {Z_\odot} {Y} \right)^{1/2} v_{200}^{5} \,
		 [F(t,t_{\rm f})]^{1/2} 
  \label{eq:sfh}
\end{equation}
where 
\begin{equation}
  F(t,t_{\rm f}) \equiv \frac {H_0^3} {v_c^{10}(t)} \int_{t_{\rm f}}^t \frac {v_c^{10}(t^\prime)} {H^{2}(t^\prime)} \, dt^\prime 
\end{equation}
and $t_{\rm f}$ is a ``formation time" at which star formation begins.

\begin{figure}[t]
\includegraphics[width=3.5in, trim = 0.0in 0.0in 0.0in 0.0in]{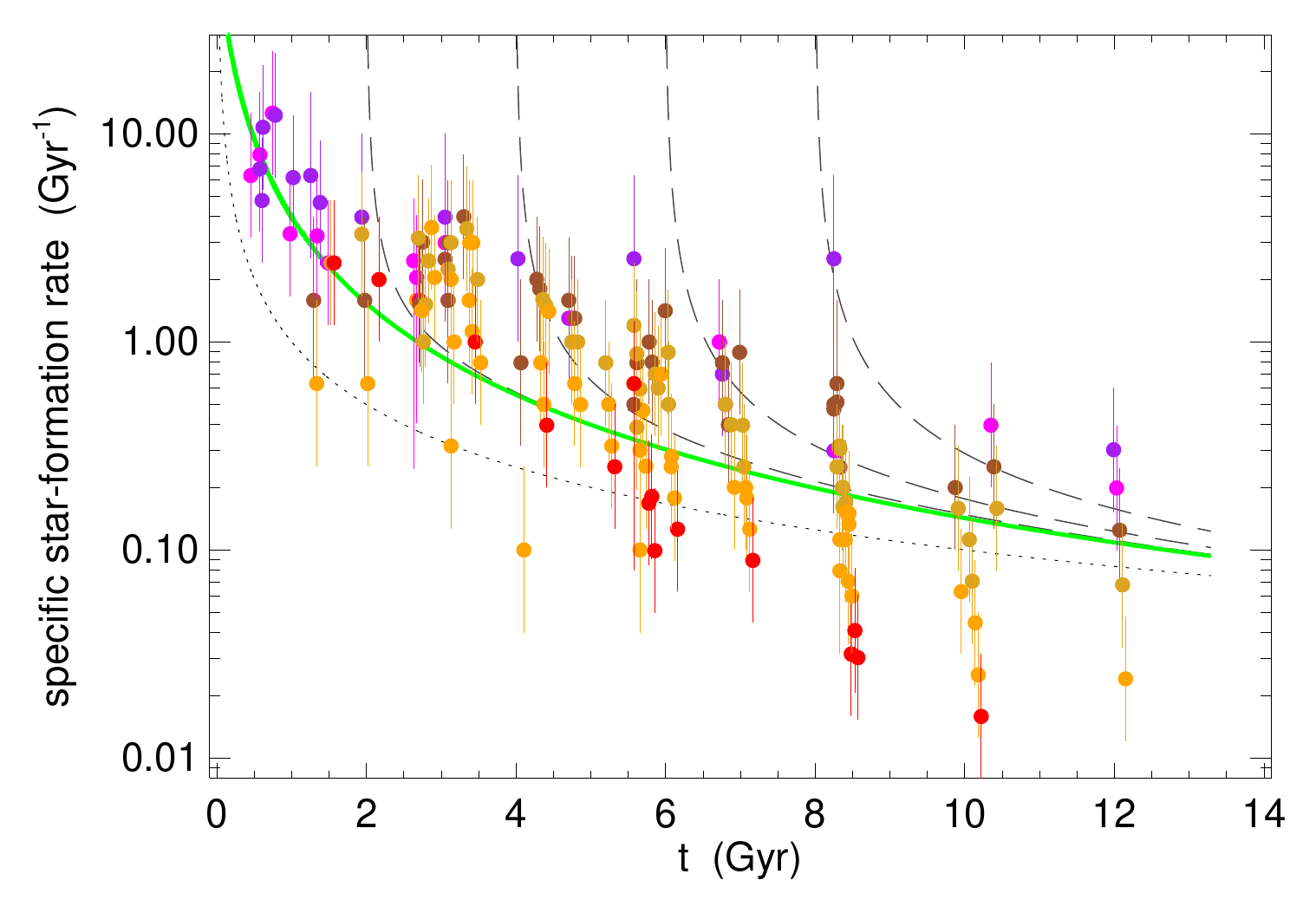} \\
\caption{ \footnotesize 
Specific star-formation rates of galaxies over cosmic time.  Points with error bars show observations of specific star-formation rates in galaxies compiled from many different sources by \citep{Behroozi+2013ApJ...770...57B}, color-coded according to galactic stellar mass:  $< 10^9 M_\odot$ (purple), $10^{9-9.5} M_\odot$ (magenta), $10^{9.5-10} M_\odot$ (brown), $10^{10-10.5} M_\odot$ (gold), $10^{10.5-11} M_\odot$  (orange),  $> 10^{11} M_\odot$ (red).  Each point represents an average over many galaxies in that stellar-mass bin at a particular epoch in cosmic time, and the error bars indicate uncertainty in the mean.  The green solid line shows specific star-formation rate predictions for $t_{\rm f} = 0$.  Dashed lines show predictions for $t_{\rm f} = 2$, 4, 6, and 8 Gyr.  The dotted line shows $t^{-1}$. 
\vspace*{1em}
\label{fig:ssfr}}
\end{figure}

Star-formation histories derived from equation~(\ref{eq:sfh}) align with measurements of specific star-formation rates in galaxies and how they trend across cosmic time (Figure~\ref{fig:ssfr}).  In this context, the precipitation threshold behaves like a ÒregulatorÓ of the sort envisioned by \citet{Lilly+2013ApJ...772..119L}.  The circumgalactic medium acts as a reservoir that collects both low-enrichment gas accreting onto the galaxy and highly enriched gas flowing out of the galaxy.  Precipitation is how gas goes from the reservoir into the galaxy, where it can form stars, and the gas transfer rate depends on $T$ and $\Lambda$.  The circumgalactic gas temperature depends in turn on the depth of the galaxy's potential well, while $\Lambda$ depends also on enrichment of the circumgalactic medium, which self-regulates through the mechanism of abundance saturation.  

\section{Enrichment and Stellar Mass} 
\label{sec-MZR}

Precipitation-regulated galaxies exhibit scaling relations among galactic stellar mass, enrichment level, and total mass that are nearly identical to the observed relations.   The mean enrichment level is $Z \approx Y M_* / (f_{\rm b} M_{500})$, which corresponds to 
\begin{equation}
  \frac {Z} {Z_\odot} \approx 0.6 \, \left( \frac {Y} {Z_\odot} \right)^{0.7} (\epsilon_* F)^{0.3} 
            \left( \frac {M_*} {10^{11} \, M_\odot} \right)^{0.4}  \frac {H} {H_0} \; \; ,
\end{equation}
according to equation (5).  The normalization of this relation is similar to observations but depends somewhat on the uncertain parameter $\epsilon_*$, which requires more detailed modeling. Instead, the scaling with $M_*$ is what provides the most compelling evidence in support of the precipitation-regulated framework.  

\begin{figure}[t]
\includegraphics[width=3.5in, trim = 0.0in 0.0in 0.0in 0.0in]{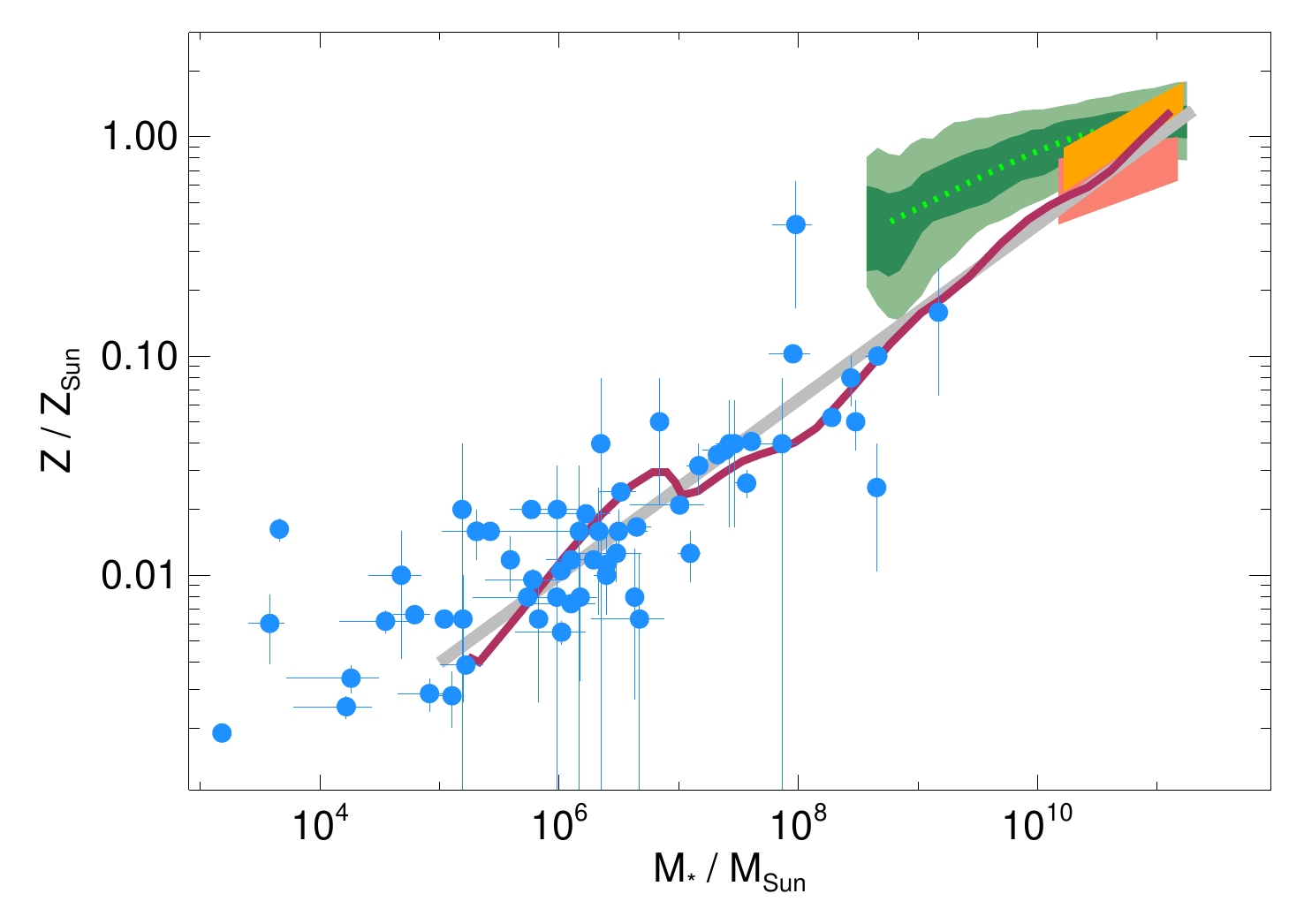} \\
\caption{ \footnotesize 
Relationship between total stellar mass $M_*$ and heavy-element abundance $Z$.  The maroon line shows how $Z(M_*)$ scales for precipitation-regulated galaxies; the grey line shows the approximation $Z \approx ( M_* /  10^{11} M_\odot )^{0.4}$. Blue points show stellar Fe abundances in low-mass galaxies from \citet{McConnachie2012AJ....144....4M}.  Orange- and salmon-colored regions show stellar Mg and Fe abundances, respectively, in elliptical galaxies from \citet{Graves+2009ApJ...693..486G}. In green is the mean gas-phase abundance derived by \citet{KewleyEllison2008ApJ...681.1183K} from data compiled by \citet{Tremonti+2004ApJ...613..898T}, along with $1\sigma$ and $2\sigma$ standard-deviation contours.
\vspace*{1em}
\label{fig:MZR}}
\end{figure}

Figure~\ref{fig:MZR} shows how that scaling corresponds to observed $Z(M_*)$ relations.  Precipitation-driven feedback naturally leads to heavy-element abundances in dwarf galaxies \citep{McConnachie2012AJ....144....4M} that are only $\sim 1$\% of the levels found in large galaxies \citep{Graves+2009ApJ...693..486G}, with little dependence on environment.  No dependence of the parameter $\epsilon_*$ on galaxy mass is required.  Radiative cooling is simply more efficient at $10^5$~K than at $10^7$~K, making the gas around smaller galaxies more susceptible to precipitation.  As a result, their circumgalactic gas densities are lower, and star formation limited by the flow of condensing clouds into the galaxy is correspondingly slower, which strongly limits the rate of heavy-element enrichment.  Furthermore, the model predicts $Z \propto M_*^{0.4} (\dot{M}_* / M_*)^{-0.3}$ for late-blooming galaxies lying above the green line in Figure~\ref{fig:ssfr}, in quantitative agreement with observations of galaxies with large specific star-formation rates \citep{Mannucci+2010MNRAS.408.2115M,Salim2014ApJ...797..126S}, since $\dot{M}_* / M_* \propto F^{-1}$ for such galaxies. 

\begin{figure}[t]
\includegraphics[width=3.5in, trim = 0.0in 0.0in 0.0in 0.0in]{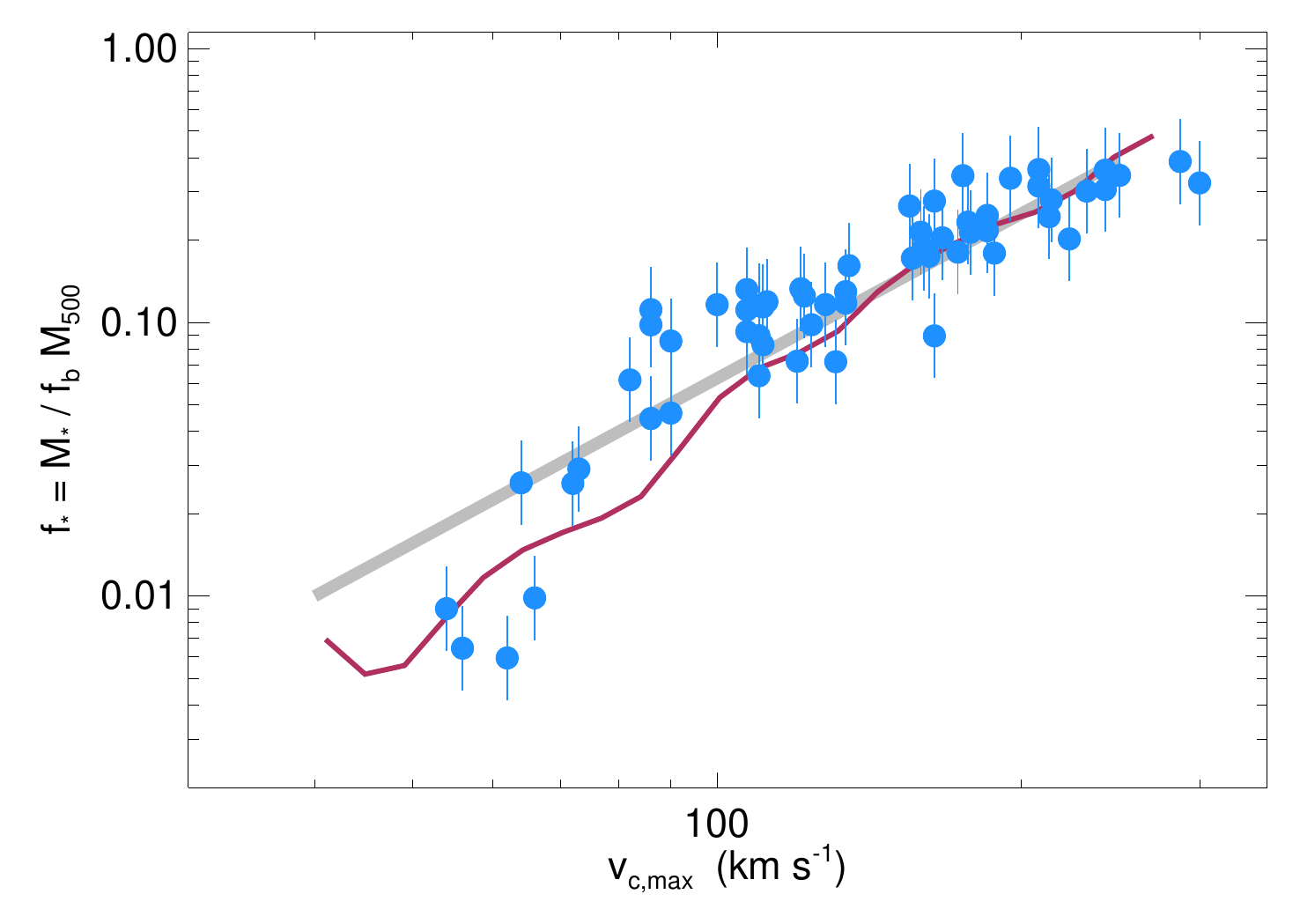} \\
\caption{ \footnotesize 
Relationship between a galaxy's stellar baryon fraction $f_*$ and the maximum circular velocity $v_{c,{\rm max}}$ of its gravitational potential.  The maroon line shows the dependence of $f_*$ on $v_{c,{\rm max}}$ for precipitation-regulated galaxies; the grey line shows the approximation $f_* \approx 0.25 \, v_{200}^2$. Points show measurements of $f_*$ in spiral galaxies from \citet{McGaugh2005ApJ...632..859M}.  
\vspace*{1em}
\label{fig:fstar}}
\end{figure}

This regulation mechanism also accounts for the observed scaling of stellar baryon fraction with $v_c$ (see Figure~\ref{fig:fstar}), predicting that
\begin{equation}
 f_* \, \equiv \, \frac {M_*} {f_{\rm b} M_{500}} \, \approx \, 0.5 \left( \epsilon_* \frac {Z_\odot} {Y} \right)^{1/2} \frac {H} {H_0} \, v_{200}^2 \; \; ,
 \label{eq:fstar}
\end{equation}
for present-day galaxies, for which $F(t,0) \approx 0.44$.  Another way to characterize this dependence is in terms of the well-known Faber-Jackson and Tully-Fisher scaling relations \citep{FaberJackson1976ApJ...204..668F,TullyFisher1977AA....54..661T}, which show that the stellar luminosity $L$ of a galaxy depends on its gravitational potential according to $L \propto v_c^\xi$, with $\xi \approx 4 - 5$.  A precipitation-regulated galaxy should have $M_* \propto v_c^5$, which gives $L \propto v_c^5$ for an invariant stellar mass-to-light ratio.  Recent observations show that the stellar mass-to-light ratio for massive galaxies depends on galactic mass, with $M_* / L \propto v_c$ \citep{Spiniello+2014MNRAS.438.1483S}, which leads to $L \propto v_c^4$ for massive galaxies and indicates that the Faber-Jackson and Tully-Fisher relations are natural outcomes of precipitation-regulated star formation.

\section{Giant Galaxies}
\label{sec-giantgals}

The trends shown in Figures \ref{fig:MZR} and \ref{fig:fstar} must change in giant galaxies for two reasons:  First, giant precipitation-regulated galaxies with deep potential wells transform their gas into stars more quickly than smaller galaxies, producing the trend in Figure~\ref{fig:fstar}, which must stop before $f_*$ reaches unity.  According to equation (\ref{eq:fstar}), galaxies with the greatest values of $v_c$ are the first to become gas-limited.  This event inevitably causes star formation to decline at a cosmic time approximately proportional to $v_c^{-4}$, possibly explaining the phenomenon astronomers call ÒdownsizingÓ \citep{MadauDickinson2014ARAA..52..415M}. Second, as the baryon balance tips from gas toward stars, black-hole feedback and supernova explosions produced by a giant galaxy's old stellar population become capable of shutting down precipitation and preventing further star formation \citep{Voit+2015ApJ...803L..21V}. Observations indicate that galaxies reach this threshold when $f_* \sim 0.3$ \citep{McGaugh+2010ApJ...708L..14M,Dai+2010ApJ...719..119D}.  According to the trend shown in Figure~\ref{fig:fstar}, galaxies with $v_c \sim 250 \, {\rm km \, s^{-1}}$ are the ones currently making this transition.  The corresponding stellar mass at the transition is $\sim 10^{11} M_\odot$ and should change with time $\propto t^{1/4}$, in accordance with the observed lack of time dependence in this transitional stellar mass \citep{Behroozi+2013ApJ...762L..31B}.

Further increases in total mass cause a qualitative change in how cosmic systems acquire stars. Once the system temperature exceeds $\sim 1$~keV, the precipitation model predicts that $f_*$ should decline with increasing system mass, because the precipitating region starts to shrink relative to $r_{500}$. Its outer boundary in this regime is given by the radius $r_{\rm precip}$ at which the electron density at the precipitation threshold equals the value $f_{\rm b} ( 2 \pi G \mu_e m_p \tff^2)^{-1}$, obtained by assuming that the electron density is proportional to the total mass density. The cooling time at this boundary in massive galaxy clusters is
\begin{eqnarray}
  \tc (r_{\rm precip}) & \; \approx \; & \frac {50} {3 \pi} \frac {f_{\rm b}} {G \mu_e m_p} \frac {\Lambda} {kT}  \\
                                & \; \approx \; & 500 \left( \frac {T} {10^8 {\rm K}} \right)^{-1/2} \, {\rm Myr} \; \; \nonumber .
\end{eqnarray}
Consequently, the mass cooling rate within the precipitation zone scales as $\dot{M}_* \propto \epsilon_* v_c^3 (r_{\rm precip})$, implying that the stellar mass of the central galaxy decouples from the ambient gas temperature and stops growing in proportion to the total mass of the system.  High-mass systems such as galaxy clusters must therefore acquire the majority of their stars by cannibalizing lower-mass systems rather than by producing new stars themselves.

\section{Central Black-Hole Mass}
\label{sec-Msigma}

\begin{figure}[t]
\includegraphics[width=3.5in, trim = 0.0in 0.0in 0.0in 0.0in]{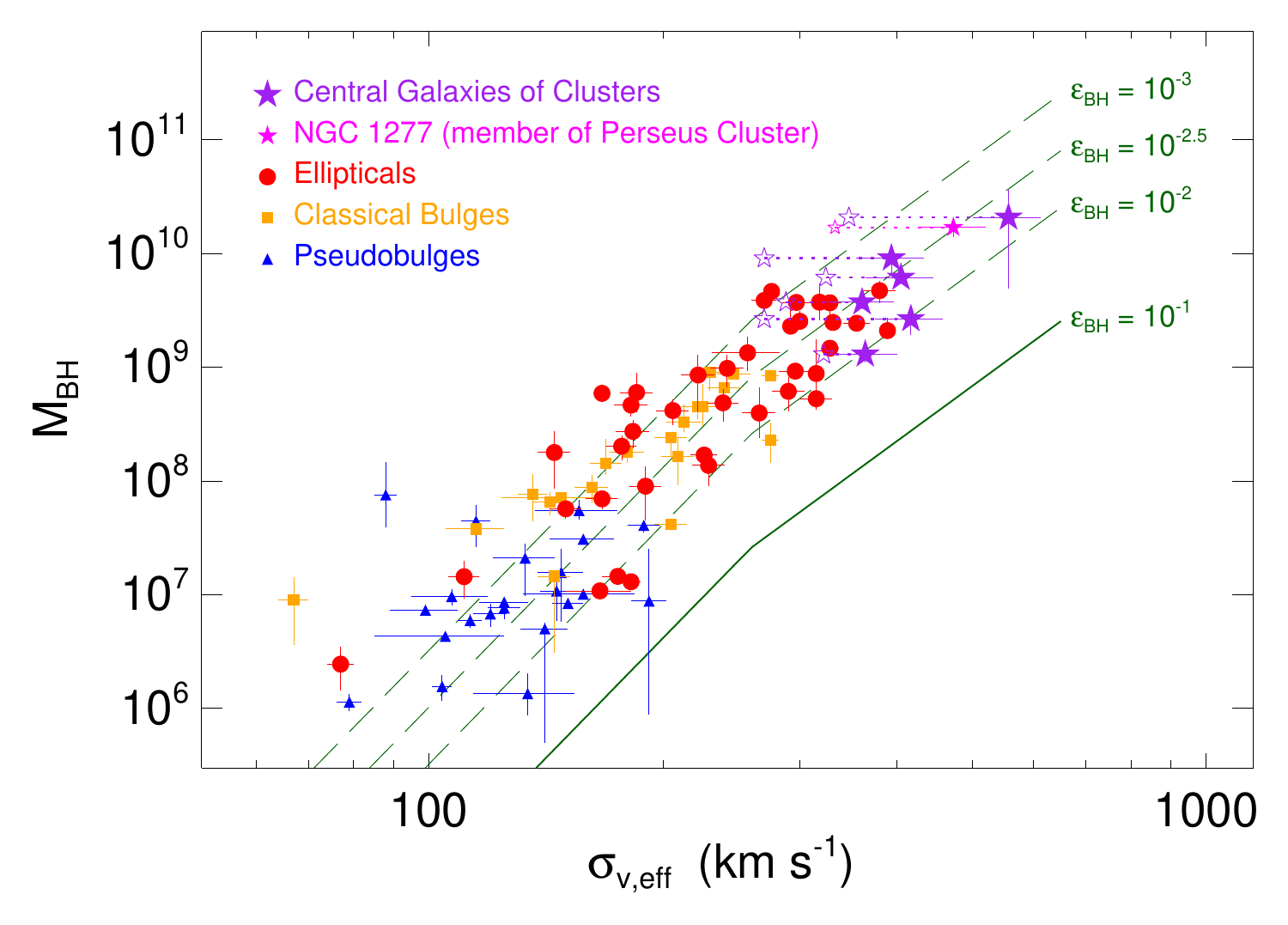} \\
\caption{ \footnotesize 
Relationship between velocity dispersion $\sigma_v$ of a galaxy's stars and the mass $M_{\rm BH}$ of its central black hole. Data points for ellipticals (circles), classical bulges (squares), and pseudobulges (triangles) are from \citet{KormendyHo2013ARAA..51..511K}.  Star symbols show how galaxies belonging to galaxy clusters shift to the right when an effective velocity dispersion $\sigma_{v,{\rm eff}} \approx \sigma_v^{3/5} [kT/(\mu m_p )]^{1/5}$ is used.  Dashed green lines show model predictions for different values of the efficiency factor $\epsilon_{\rm BH}$.  A solid green line shows the lower limit on $M_{\rm BH}$ derived for $\epsilon_{\rm BH} = 0.1$.
\vspace*{1em}
\label{fig:Msigma}}
\end{figure}

These mass-dependent features of the precipitation-regulated model appear to be reflected in the observed relationship \citep{KormendyHo2013ARAA..51..511K} between the mass $M_{\rm BH}$ of a galaxy's central black hole and the line-of-sight velocity dispersion $\sigma _v \approx \sqrt{2} \,  v_c$ of its stars.  If mass accretion onto a central black hole is the primary energy source for regulating precipitation, then a galaxy's black-hole mass $M_{\rm BH}$ should depend on the cooling rate of its precipitation zone via
\begin{eqnarray}
  \epsilon_{\rm BH} H c^2 M_{\rm BH} & \; \approx \, & \int 4 \pi r^2 n_e^2 \Lambda \, dr   \\
  	& \; \approx \; & \frac {3 f_{\rm b}} {5 G \mu_e m_p} kT \sigma_v^3 
	                \min \left[ 1 , \frac {\tff(r_{500})} {\tff(r_{\rm precip})} \right]  \nonumber
	\; \; ,
\end{eqnarray}
where $\epsilon_{\rm BH}$ is a coupling efficiency representing the fraction of accreted mass-energy that ends up as heat in the circumgalactic medium.  In massive galaxies with $\sigma_v \sim 250 \, {\rm km \, s^{-1}}$, the model predicts $M_{\rm BH} \propto \sigma_v^5$ and agrees with the data for $\epsilon_{\rm BH} \approx 10^{-3} - 10^{-2}$ (Figure~\ref{fig:Msigma}).  In precipitation-regulated galaxies with $\sigma_v < 200 \, {\rm km \, s^{-1}}$, the model predicts a steeper relation ($M_{\rm BH} \propto \sigma_v^7$), which agrees with the data for a similar coupling efficiency.  At the high-mass end, the model predicts $M_{\rm BH} \propto T \sigma_v^3$, which explains why the masses of black holes in the central galaxies of galaxy clusters tend to exceed those in other large galaxies with similar velocity dispersions.  One can account for the decoupling of central cluster galaxies from system temperature by defining an effective velocity dispersion $\sigma_{v,{\rm eff}}^5 \propto T \sigma_v^3$.  Plotting $M_{\rm BH}$ as a function of $\sigma_{v,{\rm eff}}$ instead of $\sigma_v$ then shifts central cluster galaxies onto a tighter relation corresponding to $\epsilon_{\rm BH} \approx 10^{-2.3}$.

\section{Toward Greater Realism}
\label{sec-realism}

Real galaxies are clearly more complicated than portrayed in these spartan precipitation-regulated models.  For example, the models completely ignore angular momentum, which is critical for the formation of galactic disks and is likely to shift the precipitation threshold to lower values of $\tctff$ \citep{Gaspari+2014arXiv1407.7531G}. They do not account for cold clouds that fall into galaxies without being shock-heated to the ambient gas temperature.  They do not attempt to describe the enrichment gradients observed in real galaxies. Nor do they attempt to account for environmental effects, such as stripping of the circumgalactic medium from galaxies orbiting within larger gravitationally-bound systems.

On the other hand, the models do a remarkable job of reproducing several of the most important scaling relations observed among global properties of galaxies, particularly since they rely on just a few highly plausible assumptions: (1) that galaxies acquire star-forming gas through condensation of the circumgalactic medium, (2) that condensation triggers feedback which heats the circumgalactic medium, and (3) that chemical enrichment of circumgalactic gas is proportional to the amount of star formation in the galaxy.  The precipitation threshold therefore seems likely to be an important feature of galaxy evolution, not just in the giant galaxies where it is directly observed but also in lower-mass galaxies, around which the volume-filling gas component is essentially unobservable. Such a threshold should be an emergent feature of any numerical simulation of galaxy evolution in which condensation of circumgalactic gas triggers strong feedback that suspends the ambient medium in a marginally precipitating state. If that happens, then star formation fueled by precipitating gas will lead to saturation of chemical enrichment and will produce populations of galaxies with mass-dependent trends strikingly similar to those observed among real galaxies.

\vspace*{1.0em}

GMV, BWO, and MD acknowledge NSF for support through grant AST-0908819 and NASA for support through grants NNX12AC98G and HST-AR-13261.01-A.  G.L.B. acknowledges NSF AST-1008134, AST-1210890, NASA grant NNX12AH41G, and XSEDE Computational resources. GMV is also indebted to M. Fall for a useful critique of an early draft and to A. Evrard for discussing these ideas during several long mountain-bike rides.

\bibliographystyle{apj}

\end{document}